\documentclass[apjl]{emulateapj}
\usepackage{amsmath}
\usepackage{graphicx}
\usepackage{grffile}
\usepackage{epsfig,psfrag,epic,eepic}
\usepackage{textcomp}
\usepackage{times}
\shorttitle{Constraints on the DM EoS}
\shortauthors{B.\,Sartoris et al.}

\begin{document}

\title{CLASH-VLT: Constraints on the Dark Matter Equation of State from Accurate
Measurements of Galaxy Cluster Mass
Profiles}

\author{
Barbara Sartoris\altaffilmark{1,2,3}, 
Andrea Biviano\altaffilmark{2},
Piero Rosati\altaffilmark{4}, 
% Matthias Bartelmann\altaffilmark{5}, 
Stefano Borgani\altaffilmark{1,2,3},
Keiichi Umetsu\altaffilmark{5}, 
Matthias Bartelmann\altaffilmark{6},
Marisa Girardi\altaffilmark{1,2},
Claudio Grillo\altaffilmark{7},
Doron Lemze\altaffilmark{8},
Adi Zitrin\altaffilmark{9,10},
Italo Balestra\altaffilmark{11,2},
Amata Mercurio\altaffilmark{11},
Mario Nonino\altaffilmark{2},
Marc Postman\altaffilmark{12},
Nicole Czakon\altaffilmark{5},
Larry Bradley\altaffilmark{12},
Tom Broadhurst\altaffilmark{13},
Dan Coe\altaffilmark{12},
Elinor Medezinski\altaffilmark{14},
Peter Melchior\altaffilmark{15},
Massimo Meneghetti\altaffilmark{16,17},
Julian Merten\altaffilmark{18},
Marianna Annunziatella\altaffilmark{1,2},
Narciso Benitez\altaffilmark{19},
Oliver Czoske\altaffilmark{20},
Megan Donahue\altaffilmark{21},
Stefano Ettori\altaffilmark{16,17},
Holland Ford\altaffilmark{8},
Alexander Fritz\altaffilmark{22},
Dan Kelson\altaffilmark{23},
Anton Koekemoer\altaffilmark{12},
Ulrike Kuchner\altaffilmark{20},
Marco Lombardi\altaffilmark{24},
Christian Maier\altaffilmark{20},
Leonidas A. Moustakas\altaffilmark{18},
Emiliano Munari\altaffilmark{1,2},
Valentina Presotto\altaffilmark{1,2},
Marco Scodeggio\altaffilmark{22},
Stella Seitz\altaffilmark{25,26},
Paolo Tozzi\altaffilmark{27},
Wei Zheng\altaffilmark{8},
Bodo Ziegler\altaffilmark{20}
}

\email{Barbara Sartoris sartoris@oats.inaf.it}
\altaffiltext{1}{Dipartimento di Fisica, Sezione di Astronomia, Universit\`a di Trieste, Via Tiepolo 11, I-34143 Trieste, Italy}
\altaffiltext{2}{INAF/Osservatorio Astronomico di Trieste, Via Tiepolo 11, I-34143 Trieste, Italy}
\altaffiltext{3}{INFN, Sezione di Trieste, Via Valerio 2, I-34127 Trieste, Italy}
\altaffiltext{4}{Dipartimento di Fisica e Scienze della Terra, Universita' di Ferrara, Via Saragat, 1, I-44122, Ferrara, Italy}
\altaffiltext{5}{Institute of Astronomy and Astrophysics, Academia Sinica, P. O. Box 23-141, Taipei 10617, Taiwan}
\altaffiltext{6}{Zentrum f\"ur Astronomie der Universit\"at Heidelberg, ITA, Albert-Ueberle-Str. 2, 69120 Heidelberg, Germany}
\altaffiltext{7}{Dark Cosmology Centre, Niels Bohr Institute, University of Copenhagen, Juliane Maries Vej 30, 2100 Copenhagen, Denmark}
\altaffiltext{8}{Department of Physics and Astronomy, The Johns Hopkins University, 3400 North Charles Street, Baltimore, MD 21218, USA}
\altaffiltext{9}{Cahill Center for Astronomy and Astrophysics, California Institute of Technology, MS 249-17, Pasadena, CA 91125, USA}
\altaffiltext{10}{Hubble Fellow}
\altaffiltext{11}{INAF/Osservatorio Astronomico di Capodimonte, Via Moiariello 16 I-80131 Napoli, Italy}
\altaffiltext{12}{Space Telescope Science Institute, 3700 San Martin Drive, Baltimore, MD 21218, USA}
\altaffiltext{13}{Department of Theoretical Physics, University of the Basque Country, P. O. Box 644, 48080 Bilbao, Spain}
\altaffiltext{14}{Department of Physics and Astronomy, The Johns Hopkins University, 3400 North Charles Street, Baltimore, MD 21218, USA}
\altaffiltext{15}{Department of Physics, The Ohio State University, Columbus, OH, USA}
\altaffiltext{16}{INAF/Osservatorio Astronomico di Bologna, via Ranzani 1, I-40127 Bologna, Italy}
\altaffiltext{17}{INFN, Sezione di Bologna; Via Ranzani 1, I-40127 Bologna, Italy}
% \altaffiltext{13}{Department of Astronomy, Universidad de Concepcion, Casilla 160-C, Concepcion, Chile}
\altaffiltext{18}{Jet Propulsion Laboratory, California Institute of Technology, 4800 Oak Grove Dr, Pasadena, CA 91109, USA}
\altaffiltext{19}{Instituto de Astrof\'{\i}sica de Andaluc\'{\i}a (CSIC), C/Camino Bajo de Hu\'etor 24, Granada 18008, Spain}
\altaffiltext{20}{University of Vienna, Department of Astrophysics, T\"urkenschanzstr. 17, 1180 Wien, Austria}
\altaffiltext{21}{Department of Physics and Astronomy, Michigan State University, East Lansing, MI 48824, USA}
\altaffiltext{22}{INAF/IASF-Milano, via Bassini 15, 20133 Milano, Italy}
% \altaffiltext{18}{Laboratoire AIM-Paris-Saclay, CEA/DSM-CNRS, Université Paris Diderot, Irfu/Service d'Astrophysique, CEA Saclay, Orme des Merisiers, F-91191 Gif sur Yvette, France}
% \altaffiltext{19}{Department of Astronomy, University of California, Berkeley, CA, USA}
% \altaffiltext{20}{Department of Astrophysical Sciences, Princeton University, Princeton, NJ, USA}
\altaffiltext{23}{Observatories of the Carnegie Institution of Washington, Pasadena, CA 91 101, USA}
\altaffiltext{24}{Dipartimento di Fisica, Universitá degli Studi di Milano, via Celoria 16, I-20133 Milan, Italy}
\altaffiltext{25}{University Observatory Munich, Scheinerstrasse 1, D-81679 M\"unchen, Germany}
\altaffiltext{26}{Max-Planck-Institut f\"ur extraterrestrische Physik, Postfach 1312, Giessenbachstr., D-85741 Garching, Germany}
\altaffiltext{27}{INAF/Osservatorio Astrofisico di Arcetri, Largo E. Fermi 5, 50125 Firenze, Italy}
% 
% \altaffiltext{24}{GEPI, Paris Observatory, 77 Avenue Denfert Rochereau, F-75014 Paris, France}
% \altaffiltext{25}{University Denis Diderot, 4 Rue Thomas Mann, F-75205 Paris, France}
% \altaffiltext{28}{European Laboratory for Particle Physics (CERN), CH-1211, Geneva 23, Switzerland}
% \altaffiltext{31}{CEA Saclay, Orme des Merisiers, F-91191 Gif sur Yvette, France}

% \data{}
\submitted{Submitted \today}
\begin{abstract}
A pressureless scenario for the Dark Matter (DM) fluid is a widely adopted
hypothesis, despite the absence of a direct observational evidence. According to
general relativity, the total mass-energy content of a system  shapes the
gravitational potential well, but different test particles perceive this
potential in different ways depending on their properties.
% In the weak-field limit, light sees the sum of the two Bardeen potentials, 
% while massive particles see only one of them. 
Cluster galaxy velocities, being $\ll$c, depend solely on the gravitational
potential, whereas photon trajectories reflect the contributions from the
gravitational potential plus a relativistic-pressure term that depends on the
cluster mass.

We exploit this phenomenon to constrain the Equation of State (EoS) parameter of
the fluid, primarily DM, contained in galaxy clusters. We use the complementary
information provided by the kinematic and lensing mass profiles of
the galaxy cluster MACS~1206.2-0847 at $z=0.44$, as obtained in an
extensive imaging and spectroscopic campaign within the CLASH survey. The
unprecedented high quality of our data-set and the properties of this cluster
are well suited to determine the EoS parameter of the cluster fluid.

Since baryons contribute at most $15\%$ to the total mass in clusters and their
pressure is negligible, the EoS parameter we derive describes the behavior of
the DM fluid. We obtain the most stringent constraint on the DM EoS parameter to
date, $w=(p_r+2\,p_t)/(3\,c^2\rho)=0.00\pm0.15\mathrm{(stat)}\pm0.08\mathrm{(syst)}$,
averaged over the radial range $0.5\,\mathrm{Mpc}\leq$$r$$\leq$$r_{200}$,
where $p_r$ and $p_t$ are the radial and tangential pressure,
and $\rho$ is the density. We plan to further improve our
constraint by applying the same procedure to all clusters from the ongoing
CLASH-VLT program.
\end{abstract}


%%%%%%%%%%%%%%%%%%%%%%%%%%%%%%%%%%%%%%%%%%%%%%%%%%%
\section{Introduction}
\label{sintro}
%%%%%%%%%%%%%%%%%%%%%%%%%%%%%%%%%%%%%%%%%%%%%%%%%%%
%- Why?
In the standard cosmological scenario, non-baryonic Dark Matter (DM) plays a
decisive
role, representing approximately 27\% of the mass-energy content of the Universe
\citep{planck13XVI}.
%- Description of the object
The existence and properties of weakly interacting massive DM
particles are inferred from mass measurements of galaxies and galaxy clusters
derived from kinematics, gravitational lensing or other probes. Additional
indirect
evidences come from the
influence DM has on the formation and evolution of the Large
Scale Structure of the Universe, and from its effects on the temperature
anisotropies observed in
the Cosmic Microwave Background \citep[e.g.,][]{peebles80}.

% - Assumptions, what would change
A widely adopted assumption in cosmology is that DM is a pressureless
fluid. Although such hypothesis has not been directly proven yet, in a Universe
where the DM fluid has a large pressure, there would not be enough time for
cosmological
structures to grow between the recombination epoch and today
\citep[e.g.,][]{coles02}. Moreover, the observed properties of the Large 
Scale Structure are consistent with the current 
pressureless DM scenario \citep[e.g.,][]{sanchez12,sanchez13,samushia13}.

%- Observational constraints
Despite considerable efforts for a direct detection of DM with underground experiments 
(e.g.,
DAMA\footnote{http//people.roma2.infn.it/~dama/web/home.html},
XENON\footnote{http//xenon.astro.columbia.edu/},
PICASSO\footnote{http//www.picassoexperiment.ca/},
XMASS\footnote{http//www-sk.icrr.u-tokyo.ac.jp/xmass/index-e.html},
CDMS\footnote{http://cdms.berkeley.edu/}), no significant signal has been found to date 
and 
the nature of DM remains unknown.
 
% The current pressureless scenario 
% assumes the DM Equation of State (EoS) 
% parameter to be zero.
The current pressureless scenario sets the DM Equation of State
parameter to be zero, by definition.
% - Aim of the paper
The aim of this Letter is to directly test this assumption
by constraining the EoS parameter of the galaxy cluster fluid, using
new accurate measurements of a cluster mass profile from 
lensing and kinematics analyses.

% - Previous papers 
\citet{bharadwaj03} first proposed to combine 
the analyses of the rotational curves and of the
lensing signal in spiral galaxies  to
constrain the amount of DM pressure. \citet{faber06}
generalized the \citet{bharadwaj03} approach without assuming any
model for both the DM EoS and the rotational curves. 
Finally, \citet{serra11} extended the method to the case of galaxy clusters.

% 
% In cluster the
% concentration of DM is larger than in galaxies and thus clusters are a
% stronger
% tracers of the DM EoS.

% - what's new
In this Letter, we apply this method to the $z=0.44$ cluster MACS
1206.2-0847 (hereafter MACS 1206), which has been studied
in detail as part of the Cluster Lensing And Supernova survey with Hubble
\citep[CLASH;][]{postman12}. \citet{umetsu12} have used new
high quality imaging of this cluster obtained with Subaru and the Hubble Space Telescope 
(HST), to derive
its mass density profile from weak-lensing distortion, magnification, and
strong-lensing analyses.  \citet{biviano13} (hereafter B13) have used
an unprecedented data-set of about 600 measured redshifts of cluster
members, obtained as part of a VLT/VIMOS large programme (ID 186.A-0798), to determine
the cluster mass profile over the radial range of 0.05-2.5 virial radii
by applying the Jeans equation \citep{mamon13} and the caustic
technique of \citet{dg97}.

By using the complementary information provided by the
kinematics and lensing mass profile determinations, in this Letter,
we constrain the DM EoS parameter which encapsulates information 
on the amount of fluid pressure.

%%%%%%%%%%%%%%%%%%%%%%%%%%%%%%%%%%%%%%%%%%%%%%%%%%%
\section{Theoretical Framework}
\label{stheo}
%%%%%%%%%%%%%%%%%%%%%%%%%%%%%%%%%%%%%%%%%%%%%%%%%%%
% - geometry INSIDE a cluster.
A static, spherically symmetric spacetime is described by the metric
\citep[e.g.,][]{misner73}:
\begin{equation}
  \label{eqmetric}
  \mathrm{d} s^{2}= -e^{2\Phi(r)/c^2}\, \mathrm{d} t^{2} + e^{2\lambda(r)}
\mathrm{d} r^{2} +r^{2}\,
  \mathrm{d}\theta^2 + r^{2}\,\sin^2\!\theta\mathrm{d}\varphi^2 \, ,
\end{equation}
where $\Phi(r)$ and $\lambda(r)$ are two arbitrary generic
functions of the metric.
% EoS 
One can apply to such metric the Einstein field equations, $G_{\alpha\beta} =
\left( 8\pi G/c^4\right) T_{\alpha\beta}$, by using the stress-energy
tensor of an ideal gas with radial, $T_{rr}$, and tangential,
$T_{\theta \theta}$, pressure components as $T_{\alpha\beta}=
\left(\rho+\frac{p}{c^2}\right) u_{\alpha} u_{\beta} + p
g_{\alpha\beta}$. 
In these equations, $G_{\alpha\beta}$ is the Einstein tensor, $G$
is Newton's gravitational constant, $c$ is the speed of light in vacuum,
$u_{\alpha}$ is the fluid's four velocity, and $g_{\alpha \beta}$ is the metric
tensor. 

It is then possible to
obtain the density
profile and the pressure profiles in the radial and transverse
directions \citep[e.g.,][]{schutz09}:
\begin{eqnarray}
\label{eqden_prof}
  \rho(r)\, \!&=&\! \, \frac{1}{4\pi} \frac{m^{\prime}(r)}{r^2} \, , \\
  \label{eqpr_prof}
  p_r(r) \!&=&\! - \frac{c^4}{8\pi G} \frac{2}{r^2}\, 
  \left[ 
  \frac{G m(r)}{c^2 r} - r\,\frac{\Phi^{\prime}(r)}{c^2}\left( 1-\frac{2\, G m(r)}{c^2 r}
\right)  \right] ,  \\
  \label{eqpt_prof}
  \nonumber
  p_t(r) \, \!&=&\! \, \frac{c^4}{8\pi G} \left\lbrace \left(
1-\frac{2\,G m(r)}{c^2 r} \right) 
  \frac{1}{c^2} \left[ \frac{\Phi'(r)}{r} + \frac{\Phi'(r)^2}{c^2} +\Phi''(r)
\right] \right. \\
  &~& - \left. \frac{G}{c^2} \left(\frac{m(r)}{r}\right)^\prime \left(\frac{1}{r}
+\frac{\Phi^\prime(r)}{c^2}\right)  \right\rbrace \, ,
\end{eqnarray}
where prime denotes the partial
derivative with respect to the r-coordinate, $m(r)$ is the mass within a sphere
of radius $r$, and 
the metric function $e^{2\lambda(r)}$ is defined as
\begin{equation}
\label{eqm_fun}
  e^{-2\lambda(r)} \stackrel{\Delta}{=} 1 - \frac{2\, G\, m (r) }{c^2\,r}\,. 
%   m(r) \stackrel{\Delta}{=} \frac{1}{2} \, \frac{r c^2}{G}  \left(1-e^{2
%   \Lambda}\right) \,. 
\end{equation}
%
% - Why anisotropic pressure
For a medium with isotropic pressure, 
the metric Eq. \ref{eqmetric} (describing the geometry of the spacetime inside
the cluster)
matches the
Schwarzschild metric (describing the geometry of the space time outside
the cluster) 
at $r=\tilde{r}$, where $\tilde{r}$ is
the size of the cluster,
only if $p_r=p_t=0$ \citep{bharadwaj03}.

Thus, to check the validity of the pressureless assumption, throughout this
Letter, we
consider a fluid with an anisotropic pressure and the most general definition of
EoS:
\begin{equation}
\label{eqeos}
  w(r) \,=\, \frac{p_r(r) + 2\, p_t(r)}{3\, c^2 \rho(r)} \, .
\end{equation}

The functions $\Phi(r)$ and $m(r)$ fully characterize the metric, and the
density and pressure profiles, and, consequently, the EoS.
We calculate the two metric functions by using the mass profiles of
the cluster derived from the velocity distribution of cluster galaxies
and from the gravitational lensing measurements.

In general relativity, all the mass-energy content of the cluster
shapes the gravitational potential well. However, the two probe
particles used in the aforementioned analyses, the
galaxies and the photons, perceive the gravitational potential in different
ways, due
to their distinct properties.

To calculate the trajectory of the galaxy test particles, with
velocity $v\ll c$ (B13 measured the velocity dispersion of the MACS 1206 cluster
$\sigma_{los} =
1087^{+53}_{-55} \; \mathrm{km\, s}^{-1}$ ), we relate
Einstein's field equations to the Poisson equation in the weak field
approximation $(2\Phi \ll c^2 \;\mathrm{and}\; 2mG/r\ll c^2 )$. Thus,
the (0,0) component of Einstein's field equations reads
\begin{equation}
\label{eqpoiss}
R_{00} \approx \nabla^2
\Phi = \frac{4 \pi G}{c^2} \left( c^2 \rho + p_r + 2 p_t \right) 
\end{equation}
for the cluster fluid \citep[e.g.,][]{schutz09}, where the metric
potential $\Phi$ is different from $\Phi_N$, the Newtonian potential. In the
Newtonian limit, $\rho\gg p/c^2$ and we recover the usual
Poisson equation $\nabla^2 \Phi_N = 4 \pi G \, \rho$.

From Eq. \ref{eqpoiss}, we can see that the mass profile derived from the
kinematics analysis, $m_k (r)$, depends only on the (0,0) part of the metric.
Therefore 
the relation between the Poisson equation (Eq. \ref{eqpoiss}) and this
mass profile is \begin{equation}
\label{eqmkin}
m_k (r) = \frac {r^2}{G} \nabla{\Phi}(r)\,.
\end{equation}

Since, in the lensing analysis, the probe particles travel at the
speed of light, the full treatment of the geodesics is needed, even in
the case of weak field approximation. \citet{misner73} \citep[see
also][]{faber06} derived the
expression of the lensing potential by applying Fermat's principle to
the geodesics of the photons moving through the cluster potential described by
the metric 
Eq.\ref{eqmetric}.  In this context, the light-ray trajectory is fully
described by the relativistic analogue of the refractive index. For
the metric Eq.\ref{eqmetric}, the
effective refractive index in the weak field approximation is
\begin{eqnarray}
\label{eqref_index}
\nonumber
 n(r) \, \!&=&\! \, 1 - \frac{\Phi (r)}{c^2} - \frac{G}{c^2} \int \frac{m(r)}{r^2} dr +\\
 &~& + \,O \left[ \left( \frac{2Gm}{c^2r}
\right)^2, \frac{2Gm}{r} \frac{\Phi}{c^2}, \frac{\Phi^2}{c^4} \right]
\end{eqnarray}
and it is possible to define the lensing potential as
\begin{equation}
\label{eqphilen}
 2 \Phi_l (r) = \Phi (r) + G  \int \frac{m(r)}{r^2} dr \,.
\end{equation}
The trajectory of a light particle is determined entirely by the
effective refractive index $n(r)$, thus the bending of light is not only due to
$\Phi (r)$ (see Eq. \ref{eqpoiss}), but also to an extra term due to the relativistic-pressure of the fluid.

% Thus due to the presence of an extra pressure term perceived by the light
% particle, the trajectory of the probe particles themselves is defined by the
% gravitational
% potential $\Phi (r)$ and the function m(r) of the metric, the mass of the
% cluster.

The mass profile derived from the lensing analysis can be related to the lensing
potential through the Poisson equation
\begin{equation}
\label{eqrholen}
\rho_l (r) = \frac{1}{4 \pi G} \nabla^2 \Phi_l (r)
\end{equation}
\begin{equation}
\label{eqmlen}
 m_l(r) = \frac{r^2}{G} \Phi^{\prime}_l (r) = \frac {r^2 }{2 G}
\Phi^{\prime}(r)
+ \frac{m(r)}{2} = \frac{m_k(r)}{2} + \frac{m(r)}{2} \,.
\end{equation}
Note that, while according to Eq.\ref{eqpoiss} the observed 
galaxy kinematics depends on the metric component $g_{00}$ alone, the
observed gravitational lensing potential reflects contributions from
both $g_{00}$ and $g_{rr}$.  Using the kinematic and the lensing
mass profiles, we can finally determine the two metric functions $\Phi (r)$ (from Eq.\ref{eqmkin})
and $m(r)$ (from Eq.\ref{eqmlen}) and calculate the EoS of the cluster fluid
(Eq.\ref{eqeos}).

%%%%%%%%%%%%%%%%%%%%%%%%%%%%%%%%%%%%%%%%%%%%%%%%%%%
\section{The MACS 1206 cluster Mass Profiles}
\label{sdata}
%%%%%%%%%%%%%%%%%%%%%%%%%%%%%%%%%%%%%%%%%%%%%%%%%%%

The X-ray selected MACS 1206 cluster,
at redshift 0.44, has been observed in the course of the CLASH survey. 
HST observations were completed in 2011. 
A detailed
strong lensing model, based on the identification of 50 multiple
lensed images of 13 background galaxies, was presented by
\citet{zitrin12}. The combination of the inner mass density profile
from this model with weak lensing shear and magnification measurements
from {\em Subaru} multi-band images led to a reliable determination of
the mass density profile of MACS 1206 out to $\sim 2$
Mpc\footnote{\citet{umetsu12} and B13 adopted $\Omega_m =0.3, \Omega_{\Lambda}
=0.7, h=0.7$.} \citep{umetsu12}.

The spectroscopic observations with the VLT, which led to a total of
2749 objects with reliable redshift measurements in the cluster field
and the kinematic analysis, are described in B13. Using the projected
phase-space distribution of these objects and several techniques for
the rejection of interlopers, 592 cluster members were
identified. This large spectroscopic sample was used to determine
the kinematic mass profile out to the virial radius ($\sim2$ Mpc) by solving
the Jeans equation with the MAMPOSSt \citep{mamon13}
technique, further extended to 5 Mpc with the caustic method of
\citet{dg97}. 

The kinematic determination of the cluster mass profile
is in very good agreement both with the lensing determination, and
with that based on X-ray {\em Chandra} observations, under the
assumption of hydrostatic equilibrium \citep[see][B13]{umetsu12}.

The fact that different probes of the cluster mass profile converge to
similar results\footnote{
Note that the similarity of the kinematic and lensing mass profiles does
not imply a pressureless fluid $(w(r) =0)$, since, in first
approximation, $w(r)$ does not depend on the profiles, but on their derivatives
\begin{equation}
 w(r)\approx \frac{2}{3}\frac {m_k^{\prime}(r) - m_l^{\prime}(r)}{2 m_l^{\prime}(r) -
m_k^{\prime}(r)}
\end{equation}
\citep{faber06}.
} 
suggests that systematic effects in the mass
determination are negligible and that the cluster is dynamically
relaxed.
Moreover, the analysis of \citet{lemze13} did not find significant level of
substructure
within this cluster, when using the most conservative membership selection.
% Specifically, the cluster does not appear to be
% significantly
% elongated along the line-of-sight (this would bias upward the
% kinematics- and lensing-derived mass profiles). 
The concentric distribution of different mass components (dark matter, stellar
light
and gas) also underscores an equilibrium configuration. All these
properties make MACS 1206 an ideal candidate for testing the EoS of the cluster
fluid. 

\citet{umetsu12} parametrized the lensing mass profile of MACS 1206 with 
the NFW model \citep{navarro97}. The same model was found by B13 to provide the
highest likelihood
fit to the kinematic data. We therefore use the NFW model parametrization of the
kinematic and lensing mass profiles in our analyses.
To check the sensitivity of our results on the kinematic mass profile used, we
also consider 
the \citet{hernquist90} and \citet{burkert95} profiles. These models
are quite different from NFW and yet were found to provide acceptable fits to
the kinematic data (B13). 

Unlike the lensing determination of the cluster mass profile, the
kinematic determination also requires modeling the velocity
anisotropy profile of the tracers of the gravitational potential,
$\beta(r)$.  B13 considered three possible ansatz models, named 'O',
\begin{equation}
 \beta_{\mathrm{O}}(r) \,=\, \beta_{\infty} \frac{r-r_{-2}}{r+r_{-2}} \, ,
\label{eq:betao}
\end{equation}
'T' \citep[from][]{tiret07},
\begin{equation}
\beta_{\mathrm{T}}(r) \,=\, \beta_{\infty} \frac{r}{r+r_{-2}} \, , 
\label{eq:betat}
\end{equation}
and 'C' with a constant anisotropy with radius.
In Eqs. \ref{eq:betao} and \ref{eq:betat}, $r_{-2}$ is the 
scale radius at which the logarithmic derivative of the mass density
profile equals $-2$, $d \,\ln \rho /d\, \ln r = -2$ 
and $\beta_{\infty}$ is the anisotropy
value at large radii. The O model gives the
smallest product of relative errors in the two free parameters of the
mass profile ($r_{200}$\footnote{The radius $r_{200}$ is the radius of a sphere with
mass overdensity $\Delta$=200 times the critical density of the Universe at
the cluster redshift. We use $r_{200} \simeq 2$ Mpc from \citet{umetsu12} and
    B13.} and
$r_{-2}$),  
namely it maximizes
the ratio $(r_{200} \,  r_{-2})/(\delta r_{200} \, \delta r_{-2})$, where $\delta r_{200}$
and $\delta r_{-2}$ are the (symmetrized) errors on  $r_{200}$ and $r_{-2}$, respectively.
Therefore the NFW+O model is adopted as our
reference model in the following analysis.

%%%%%%%%%%%%%%%%%%%%%%%%%%%%%%%%%%%%%%%%%%%%%%%%%%%
\section{Results}
%%%%%%%%%%%%%%%%%%%%%%%%%%%%%%%%%%%%%%%%%%%%%%%%%%%

In this section, we present and discuss our results on the EoS of the
DM fluid as obtained from the analysis of the lensing and kinematic
mass profiles of MACS 1206 described above.  Derivatives of the
potential and the mass profiles in
Eqs.(\ref{eqden_prof}-\ref{eqpt_prof}) are computed directly from the
models of \citet{umetsu12} and B13. For details see Section 2 and 3.
% We use in Eq.\ref{eqmlen} the mass profile as obtained by \citet{umetsu12}
% from the lensing analysis using the NFW prescription, and we use the
% results 
% from B13 to calculate the kinematic mass profile,
% considering as reference model the NFW 
% one with the $\beta=\mathrm{O}$ prescription for the anisotropy parameter. 

\begin{figure}[t]
 \begin{minipage}[htb]{0.48\textwidth}
   \centering
   \includegraphics[width=0.68\textwidth,angle=270]{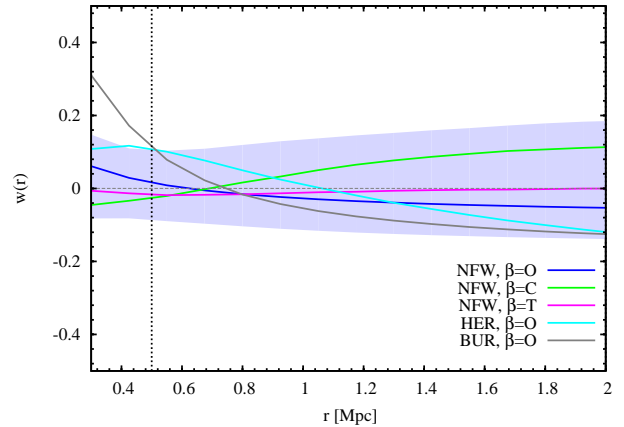}
 \end{minipage}
    \caption{EoS parameter of the cluster fluid as a function of cluster centric
radius obtained by using different prescriptions for the mass and the anisotropy
profiles ($\beta$) in the kinematic analysis: NFW $\beta$ =O (solid blue curve), NFW
$\beta$ =C (solid green curve), NFW $\beta$ =T (solid magenta curve), Burkert
$\beta$ =O (solid grey curve), Hernquist $\beta$ =O (solid cyan curve). The
shaded area represents the errors for our reference model (NFW+O), calculated
through a MonteCarlo sampling method. Errors on the other models are not shown
for the sake of clarity. The vertical dotted line indicates the lower radial
limit above which mean $w$ values in Table \ref{t:w} are computed.
}
\label{figeos_profiles}
 \end{figure}

In Fig.\ref{figeos_profiles}, we show the resulting EoS parameter
$w(r)$ (Eq.\ref{eqeos}) as a function of the cluster centric
radius. The statistical errors have been calculated via $10^4$
MonteCarlo resamplings  by propagating the uncertainties on the
parameters of the lensing and kinematic mass profile models, $r_{200}$ and
$r_{-2}$.
As for the parameters derived from kinematics analysis, B13 shows that $r_{200}$
and $r_{-2}$ have
uncorrelated errors thus we use the probability distributions shown in
Fig. 9 of B13 to explore the parameter space in the MonteCarlo sampling.
\citet{umetsu12} shows that the joint weak and strong lensing contours of the
mass profile model are elliptical, so we can assume that they are Gaussian
distributed but with covariance between the model parameters. 
We perform a MonteCarlo Markov Chain fitting to the weak and strong
lensing radial profiles to generate posterior probability distribution functions
in $r_{200}$ and
$r_{-2}$ to consider the covariance between these two parameters in the calculation
of the errors on the EoS parameter profile.

In
Fig.\ref{figeos_profiles}, the shaded area indicates the error computed
from the $16^{th}$ and $84^{th}$ percentiles of the probability
distribution at varying radii, for the reference NFW+O model. In Table
\ref{t:w}, we list the
mean values of $w$, calculated over the radial range $0.5\,
\mathrm{Mpc} \leq r\leq
r_{200}$, 
 where $w(r)$ is approximately constant. 
Since the errors on the EoS parameter at different radii, $w(r)$, are highly correlated, we 
list $w$ errors in Table \ref{t:w} obtained by computing the average of the $w$ 
uncertainties over the considered radial range. 
We limit the radial range to radii
  $\leq r_{200}$ because at larger radii both the kinematics and the
  lensing determinations of the cluster mass profile might be affected
  by systematics (deviation from dynamical relaxation for the
  kinematics, and contamination by a large-scale structure filament
  for the lensing). Moreover, the DM, whose EoS we want to constrain, dominates the mass budget in
this radial range.

Both in the Figure and in the Table, we show how our results are modified when
we use different models for the
anisotropy and kinematic mass profiles.
% Moreover, we check the dependence of our results on the mass profile model
% adopted in the
% kinematic analysis comparing the $w$ values obtained by using the NFW model
% with
% those calculated assuming the 
The results on $w(r)$ are sensitive to the adopted models, but still
within the statistical uncertainties of the reference models.

In Table \ref{t:w}, all the mean values are consistent with zero, thus with the
usually
adopted pressureless assumption. 
Note that the mean values cannot be directly inferred from the curves of
Fig.\ref{figeos_profiles} because the errors are asymmetric. 
% As for the results obtained
% with the Hernquist and the Burkert prescription 
% the values for $w$ are inconsistent with the zero values, and the errors are
% much lower reflecting the smaller relative errors on the mass profile
% parameters
% (see Table 2 in B13). 
In particular, for the reference NFW+O model,
we find a radially averaged value
\begin{equation}
w = 0.00\pm 0.15 \rm{(stat)}
\pm 0.08 \rm{(syst)} \, , 
\end{equation}
where the statistical error is listed in Tab. \ref{t:w} and
the systematic error reflects the peak to peak variation in the mean values of
$w$ obtained using different mass and 
anisotropy models.

In our estimate, we cannot disentangle the mass profile of the
baryons from that of the DM, however baryons contribute at most $15\%$ of the
cluster total mass at all radii \citep{biviano06}.
Moreover,  we estimate that the baryon contribution to $w$ is $\sim 10^{-5}$.
This is found by using the EoS of an ideal gas for the hot intra-cluster medium.
We can therefore 
assume that the EoS parameter found in our study describes the
behavior of the DM fluid in the cluster.

A previous analysis of the DM EoS \citep{serra11} has found tentative evidence
for a negative value of $w(r)$. This result was obtained by using the lensing
and kinematic mass determinations of two clusters: Coma and CL0024+1654, 
which however are known to contain major substructures
\citep{adami05,czoske02}, and their presence could affect the kinematic mass
profile determinations. 
In particular, CL0024+1654 is possibly in a post-merger state after the core
passage of two clusters occurring along the line of sight \citep{czoske02,umetsu10}.
In any case, our constraints on $w$ are significantly
more precise (by a factor of $\gtrsim 3$ than those obtained by \citet{serra11}
thanks to
the significantly better quality of both our lensing and kinematic data.

\begin{table} 
\centering
\caption{Mean values and errors for the EoS parameter of the cluster fluid}
\begin{tabular}{|l|c|r|c|}
\hline 
 & & &  \\
Profiles & $\beta$ & w$\;\;\;\;$ & $\Delta w$ \\
 & & &  \\
\hline 
 & & &  \\
\textbf{NFW}& \textbf{O}& $\textbf{0.00}$ &\textbf{0.15} \\
NFW& C&  0.15 &0.19 \\
NFW& T&  0.08 &0.18 \\
BUR& O& $-0.01$ &0.15 \\
HER& O& $0.06$ &0.19 \\
&&&\\
\hline
\end{tabular}

\bigskip

\begin{flushleft}
Notes: The mean values are obtained within the radial range $0.5\,\mathrm{Mpc}\leq r
\leq r_{200}$, using different prescriptions of the mass profile
and anisotropy parameter in the kinematic analysis. Column 1: Mass
density profiles. Column 2: anisotropy parameter. Column 3: mean values for the
EoS parameter. Column 4: average of the errors on $w(r)$ over the considered radial range.
\end{flushleft}

\label{t:w}
\end{table}

%%%%%%%%%%%%%%%%%%%%%%%%%%%%%%%%%%%%%%%%%%%%%%%%%%%
\section{Summary and Conclusions}
%%%%%%%%%%%%%%%%%%%%%%%%%%%%%%%%%%%%%%%%%%%%%%%%%%%
% In general relativity we have to accomplish 
% all the mass-energy contribution in shaping the potential well, 
% and in our case both the density 
% and the pressure give their contribution to the potential well. 
% The way in which
% the probe particle probe the underling potential fields depends 
% on the characteristic on the particle itself.
% 
% At the level of precision that we can reach with the MACS 1206 data
% from the CLASH project, it is critical to reduce the cluster-to-cluster
% systematics in
% the calculation of the mass profile. This will be possible by using
% the complete sample of 12 CLASH clusters of the VLT spectroscopic
% program.
% 
In this Letter, we discussed how the pressureless assumption for the
DM fluid can be quantitatively verified and we obtained the most
stringent constraint on the DM EoS available to date by using high quality
kinematic
and lensing mass analyses of the relaxed CLASH cluster MACS 1206.

We confirmed the pressureless assumption, namely $w = 0.00\pm 0.15 \rm{(stat)}
\pm 0.08 \rm{(syst)}$. We find no radial dependence of $w$ outside the central (0.5 Mpc)  region.

The CLASH ­ VLT final sample will provide accurate mass profiles for
12 clusters, allowing us to place stronger constraints on the DM EoS
parameter by stacking the information from all clusters. This will reduce our 
statistical errors, and,
most importantly, possible systematic effects in the mass profile
determinations such as
departure from dynamical equilibrium, and contamination by
large scale structure along the line of sight.

If a departure from sphericity of the cluster
potential well is detected, it is still possible, in principle, to apply the
method used in this Letter to calculate the DM EoS parameter. \citet{faber06}
showed how, given a mass distribution,
it is always possible to recover the corresponding density and pressure
distributions in absence of any particular potential symmetry and thus calculate
the DM EoS parameter.
From an observational point of view, cluster orientation and
asphericity can systematically affect the mass profile determinations and
consequently the EoS parameter. It is possible to reduce the impact of these
errors by stacking results derived from a large sample of clusters.

\vspace*{-5mm}

\bigskip

\begin{acknowledgements}
We would like to thank the Referee,  Michael Strauss, for his constructive and thoughtful
comments.
BS thanks Marino Mezzetti and Pierluigi Monaco for useful discussions.
This work has been partially supported by the PRIN-MIUR09 ``Tracing the growth
of 
structures in the Universe'', by the PD51 INFN grant and by the PRIN INAF 2010
``Architecture and Tomography of Galaxy Clusters''.
PR acknowledge partial support by the DFG
cluster of excellence Origin and Structure of the Universe
(http://www.universe-cluster.de). 
Support for AZ was provided by NASA through Hubble Fellowship grant
HST-HF-51334.01-A awarded by STScI.
 AF acknowledges the support by INAF through PRIN 2008 (VIPERS) and
    PRIN 2010 (VIPERS) grants.
This research was carried out in part at the Jet Propulsion Laboratory,
California Institute of Technology, under a contract with NASA.
\end{acknowledgements}

\bigskip
\bigskip
\bigskip
\bibliographystyle{apj}
% \bibliography{dark}

\clearpage

\end{document}